# MODELING OF SUBSURFACE ICE MANTLE ON INTERSTELLAR DUST GRAINS WITH ASTROCHEMICAL CODE ALCHEMIC (RESEARCH NOTE)

**Juris Kalvāns**

Engineering Research Institute "Ventspils International Radio Astronomy Center" of Ventspils University College, Inzenieru 101 a, LV-3600, Ventspils, Latvia
e-mail: juris.kalvans@venta.lv

**Abstract.** Interstellar ices are layers of molecules deposited on fine dust grains in dark and dense molecular cloud cores. Subsurface ice has been considered in a few astrochemical models, which have shown that it can be of great importance.
The aim of this work is to introduce an established subsurface ice description into the state-of-the-art astrochemical model ALCHEMIC. The model has been developed by the Heidelberg astrochemistry group. The result is an up-to-date model for interstellar molecular cloud research with possible application for protoplanetary disks.

**Key words:** astrochemistry – molecular processes – interstellar medium: clouds, dust, molecules

## 1. Introduction

### 1.1. Ices in interstellar clouds

Molecular clouds constitute approximately half of the gas mass in the Galaxy. They undergo fragmentation and gravitational contraction during their evolution. The fragments develop dense and dark cloud cores that are the birth sites of stars in the Universe.

There are several sources of energy in the interstellar medium; interstellar ultraviolet radiation field, cosmic rays (CR), cosmic-ray-induced photons and exothermic reactions are relevant for quiescent molecular clouds. The source of the interstellar radiation that permeates the interstellar medium is hot stars. The dark cores are almost completely shielded from their photons. Cosmic rays are high-energy atomic nuclei that mostly arise in interstellar shocks produced by supernovas (Aharonian et al. 2012). They penetrate even into the most dense cloud cores (Dalgarno 2006). Cosmic rays and secondary electrons excite $H_2$ molecules in clouds, which produces an internal UV radiation field within the clouds (Prasad & Tarafdar 1983). The dissociation of molecules produces chemical radicals (e.g. free atoms, OH, $CH_3$, etc.), which in their turn undergo exothermic reactions, both, in gas and solid phases (on dust grains).

The elemental composition of the interstellar medium is approximately 90 % H, 9 % He and 1-2 % heavier elements. Elements heavier than He are loosely referred to as metals. In the clouds, H is in the form of $H_2$. Refractory metal species constitute the dust. Dust may contain approximately half of metal mass. Molecular cloud cores are dark (extinction of the interstellar radiation field $A_V > 3$) and dense (density of hydrogen nuclei $n_H \geq 10^3$ cm$^{-3}$). Volatile species begin to accumulate on the surfaces of the dust grains. They form a layer of ice, major components of which include water, CO, $CO_2$, and methanol (Öberg et al. 2011). Grain surface and the icy layer are the sites of solid-phase chemistry in interstellar clouds.

The molecules in interstellar clouds are subjected to chemical and physical transformations induced by photons and CRs. They undergo accretion on grains, migration, binary reactions on the surface, photodissociation, and desorption back into the gas phase. The transformations of ice are not limited to the surface. The whole ice layer is subjected to chemical and structural changes due to CRs, CR-induced photons, and exothermic reactions (e.g. Palumbo et al. 2010; Öberg et al. 2010; Accolla et al. 2011).

### 1.2 Model of interstellar ice mantle

In our previous papers (Kalvāns & Shmeld 2010, 2013, from now on KS2013) a gas-grain astrochemical model for quiescent molecular cloud cores was developed. The model considered the subsurface ice (mantle) phase, which is omitted in many current astrochemical models. It is, perhaps, the first and currently most complete model that attempts to describe the chemical transformation of the subsurface ice. This was done via the concept of closed cavities in the mantle. They are reactive surfaces, isolated from the outer ice surface.

The current model, however, has some drawbacks that have to be solved for better quality results: first, it does not describe the attenuation of interstellar photons properly, even though this is unimportant for molecular cloud cores. Charge exchange between gas and dust particles, anion reactions, stochastic effects for surface reactions are not included in the model either. Besides, the reaction database used (Hasegawa et al. 1992 and Hasegawa & Herbst 1993) is outdated and fails to include several important reactions, particularly, hydrogenation of formaldehyde.

To solve these problems, it was decided to incorporate the mantle-phase chemistry in an existing up-to-date astrochemical model, and not to work on the further development of the program used for Kalvāns & Shmeld (2010, 2013). The model "ALCHEMIC", developed by the Heidelberg astrochemistry group was chosen as the basis for further research. The code was kindly provided by Dmitry Semenov. "ALCHEMIC" utilizes the Ohio State University reaction database[1] (version 2008_03). It includes an extended set of surface reactions, relative to that of Hasegawa et al. (1992); Hasegawa & Herbst (1993). Stochastic effects for surface reactions can be taken into account with the modified rate equation approach (Caselli et al. 1998).

In addition to fixing the above mentioned deficiencies, "ALCHEMIC" will allow to model ice chemistry in protoplanetary disks. Full temperature dependence is included for relevant transformations and reactions of the ice phase. Because of the swap of reaction databases, the new model currently does not include deuterium chemistry. This was an important part of the KS2013 paper. The new model is dubbed "Alchemic-Venta".

## 2. Rate calculation for processes implemented in "Alchemic-Venta"

The ice mantle model has been explicitly described in KS2013. "ALCHEMIC" is described by Semenov et al. (2010, S2010+). The focus of this paper is on the changes made during implementation of mantle description into the "ALCHEMIC" and their consequences. The

1 Available at Eric Herbst's home page, http://www.physics.ohiostate.edu/eric/research.html

models themselves are not described in detail. In the new model, only processes regarding ice formation and processing have been supplemented or changed.

Some of the changes arise from the fact that the model by Kalvans & Shmeld is designed for cloud modeling, while that of Semenov et al. is mainly for protoplanetary disks. The 'low metal' elemental abundances used by S2010+ were replaced with those used in KS2013, because the latter considers the whole ice mantle, not just its surface. The elements Si, P, and Cl were not considered in KS2013. Their abundances were taken from Jenkins (2009) with $F_* =$ 1 for "Alchemic-Venta".

## 2.1. Accretion

The accretion of species onto grains is significantly faster in KS2013 than in S2010+. Calculations at integration time $t$ = 1 Myr with temperature $T$ = 10 K and density $n_H = 2\times10^4$ $cm^{-3}$ result in an ice-to-gas abundance ratio of chemically active metal elements C, N, O, and S, which exceeds 100:1 for the former and is approximately 2:1 for the latter model. This is inconsistent with an earlier research, where metal depletion onto grains is expected to occur on the order of $10^5$ yr at densities similar to $10^4$ $cm^{-3}$ (e.g. Leger 1983).

In the new model the rapid accretion approach was used, as outlined by Willacy & Williams (1993). It results in a much higher rate coefficient. It also takes into account molecule accumulation onto small grains. This allows easy comparison of the results from "Alchemic-Venta" with those from KS2013, because the rapid accretion approach is the one we used in earlier papers.

All the models (Kalvāns & Shmeld 2010; Semenov et al. 2010, and "Alchemic-Venta") consider larger 0.1 μm grains in their physical and chemical description. However, molecule accretion on small grains does not contradict this approach, because the smaller grains coagulate and/or stick onto the larger grains (Köhler et al. 2012).

Positive ions stick to negatively-charged grains some 18 times more efficiently than neutral species (Umebayashi & Nakano 1980; Willacy & Williams 1993). This phenomenon is not described in the original "ALCHEMIC" code. A direct implementation of this process into the model, with the grains neutralized upon each accretion event, has resulted in mostly neutral grains. This seriously alters the ionization fraction. Charge balance may profoundly affect the chemical composition of the gas phase (Kalvāns & Shmeld 2012).

It has been decided not to include this process in the current simulations, in line with S2010+ and KS2013, until a more detailed investigation has been made. The interaction between electrons and grains has been retained 'as is' in "ALCHEMIC".

## 2.2. Desorption

All desorption mechanisms considered in KS2013 were transferred to "Alchemic-Venta". Evaporation and desorption due to whole-grain heating by direct CR hits was retained exactly as described in S2010+. Desorption by the interstellar radiation field and by CR-induced photons is included, following Roberts et al. (2007). The Habing field ($10^8$ $cm^{-2}s^{-1}$) is taken as the flux of interstellar UV photons (this mechanism was not included in KS2013). Desorption by the energy released from the reaction H + H on grain surfaces was included in line with Roberts et al. (2007), too. The only mechanism that directly transports subsurface mantle molecules into the gas is impulsive ejection of molecules by cosmic rays passing through the grain (Johnson et al. 1991). For the sake of completeness, it was included in both, surface and mantle phases in "Alchemic-Venta". In KS2013 it was considered for the mantle phase, only. It should be noted that, according to the current understanding, the yield (or efficiency) for all non-thermal desorption mechanisms is unknown, at least within an order of magnitude (see e.g. Roberts et al. 2007).

Selective desorption of important mantle species greatly affects the abundance of elements in gas and solid phases. Table 1 shows that the ice-to-gas abundance ratio for different elements may differ by orders of magnitude. This ratio is dependent on the ability for an element to storage itself into simple and sticky molecules, such as $H_2O$ or SiO. The atoms of the metallic elements have very high desorption energies.

**Table 1.** *Total ice-to-gas abundance ratio for metal elements after integration time t = 2 Myr.*

| Element | C | N | O | Si | S |
|---|---|---|---|---|---|
| ice/gas | 6.00E+01 | 7.50E+00 | 2.30E+02 | 3.30E+04 | 6.40E+02 |
| Element | Fe | Na | Mg | Cl | P |
| ice/gas | 9.60E+03 | 7.00E+03 | 2.50E+04 | 3.60E+04 | 2.00E+03 |

## 2.3 Transitions solid species between phases

The exchange of intact metal molecules between the three ice phases (surface, mantle and cavities) consist of six transitions in total. This is an extension over KS2013, where direct surface-cavity transition was not considered. The relative rates of these processes determine the structure of the mantle. It is characterized by (1) the outer surface area or porosity (Sect. 3.4. in KS2013) and (2) the amount of reactive species within the mantle (on surfaces of closed cavities, Sect. 4.1., KS2013). The desired mantle-to-surface (*M/S*) and mantle-to-cavity average abundance ratios both are assumed 100:1.

The rate coefficients for transitions between the ice phases were calculated as specified in KS2013, Eqs. 16-18. They are assumed to be bound to the rate of the energetic cosmic-ray iron nuclei passing through the grains. In "Alchemic-Venta", the rate coefficients were taken to be 10 times higher. This produces more rapid transition of surface species into the mantle and cavity phases. Physically, this means that the mantle is compacted from porous to dense structure 10 times faster. This is more consistent with the recent experimental results, which revealed that water ice formed on surface at 10 K is compact, not porous (e.g. Linnartz et al. 2011).

With these coefficients $M/S \approx 100 : 1$ is reached in an integration time of approximately 2 Myr. *M/S* may slightly differ between the elements.

## 2.4 Binary reactions on grains

The calculation of the rate of binary reactions was retained exactly as described in S2010+. The improvements over KS2013 include an extended reaction set, which can be used with higher or lower molecule binding energies. The modified rate equation approach for surface reactions is an interesting and valuable addition because of its possible application in calculating reaction rates for cavity reactions. Cavity surface reactions themselves were implemented into "Alchemic-Venta" in line with Kalvāns & Shmeld (2010). The relevant parameters are given by KS2013.

## 2.5 Photodissociation of molecules in ice

The "ALCHEMIC" reaction file includes a set of photoreactions for surface molecules. Both, interstellar and CR-induced photons are considered.

These photoreactions were applied to mantle and cavity species, too. The dissociation yields are provided in the Table 2 of KS2013. It is assumed that H and $H_2$, split off from surface species, escape into the gas phase (Hartquist & Williams 1990; Kalvāns & Shmeld 2010). They include photodissociation of cavity molecules and mantle molecules. The products of the latter transit to the cavity phase with 1 % efficiency. In addition, the opposite process was introduced in "Alchemic-Venta" – the transition of cavity molecule dissociation products into the inert mantle phase. The efficiency of this process was initially assumed 10 % of the 'traditional' dissociation in cavities. This process was introduced to reflect the trapping of radicals in ice (Öberg et al. 2010).

Photodissociation may initiate the migration of radicals along the depth of the mantle over several monolayers (Andersson & van Dishoeck 2008). Because of this, a bi-directional transition between the phases for the dissociation products in outer-surface and mantle phases was introduced here. It was assumed that the transition of UV or CR-induced photodissociation products from surface to mantle phases, or vice versa, occurs with 1 % efficiency relative to dissociation of molecules in the respective phases without a change of phase.

The real efficiencies of the dissociative reactions with transition between phases are unknown, and can be adjusted if such a need arises. The transition between outer-surface and cavity phases was not considered, because cavities, by definition, are isolated from the surface by at least several monolayers. Molecule exchange directly between cavities and surface can be induced by more energetic events, only (see above). They can be cosmic-ray hits or decay of radioactive nuclei in the grain.

## 2.6 The diffusion of hydrogen

H and $H_2$ have a different and more effective phase transition mechanism than other species – diffusion through the lattice of the ice. This process was included in the manner outlined by Kalvāns & Shmeld (2010, 2013).

A simple temperature dependence for hydrogen diffusion coefficient $D$ ($cm^2 s^{-1}$) was introduced, with an Arrhenius equation of type:

$$D = D_0 e^{\frac{E_{act}}{T}}, \qquad (1)$$

where $E_{act}$ (K) is the activation energy for diffusion (Strauss et al. 1994). The parameters $D_0$ and $E_{act}$ for $H_2$ were calculated from data supplied by Strauss et al. (1994) (at 25 and 43 K; Kalvāns & Shmeld 2010). For H the diffusion coefficients at 10 K (Awad et al. 2005) and 20 K (Strauss et al. 1994) were used. Such a single-equation temperature dependence with interpolation or extrapolation of limited data sets does not seem to be entirely correct. However, it makes the model more versatile with possible calculations for slightly elevated grain temperature. With this addition, now all relevant processes in the new model are temperature dependent.

There are four ice-phase transition processes in “Alchemic-Venta”. First, the CR-induced transition of intact molecules between all three phases is the dominant one. The second process is CR-induced photodissociation that connects the phases according to sequence: surface ↔ mantle ↔ cavities. Molecule dissociation by UV photons works in a similar manner but is much less important because of extinction of the interstellar radiation field in the clouds. The final phase transition is the diffusion of H and $H_2$. This mechanism is important for determining mantle composition, however, it affects free hydrogen, only.

# 3. Results

An updated, comprehensive astrochemical model has been created by the fusion of the idea proposed by Kalvāns & Shmeld and the code provided by Semenov et al. (Heidelberg astrochemistry group). The new model “Alchemic-Venta” has some extensions and improvements over the model used in KS2013. They include temperature dependence in all the processes, and extended surface reaction database. These have to be utilized and evaluated during a further research. The calculation time on a modest workstation PC (Intel Core Duo 2.33 GHz CPU, 2 GB RAM) for “ALCHEMIC” is approximately 7 seconds, but “Alchemic-Venta” typically runs for 1-2 minutes. The model has no deuterium reactions within its current database.

**Table 2.** *Comparison of calculated abundances with observational data. The final (3 Myr) result calculated in Kalvāns & Shmeld (2013), observational results of a high-mass protostar W33A and cloud core Elias 16 are compared to the new results of “Alchemic-Venta” (integration time 2 Myr). The latter results are presented with molecule adsorption energies* $E_b$ *given as fraction of the desorption energy* $E_D$.

| | KS13 | 0.33 $E_D$ | 0.33 $E_D$, MRE[a] | 0.77$E_D$ | W33A[b] | Elias 16[b] | Low-mass protostars[c] |
|---|---|---|---|---|---|---|---|
| $H_2O$ | 100 | 100 | 100 | 100 | 100 | 100 | 100 |
| $CO_2$ | 112 | 29 | 31 | 7 | 13 | 18 | 29 |
| CO | 7 | 7 | 8 | 20 | 8 | 25 | 29 |
| $CH_3OH$ | ~ 0 | 19 | 19 | 17 | 18 | < 3 | 5 |

[a] Modified rate equation approach

[b] Chang et al. (2007)

[c] Öberg et al. (2011)

A short glimpse of the initial results (Table 2) with the given input parameters looks very promising. $H_2O$, $CO_2$, and CO have mantle abundances in proportions comparable to interstellar ices (e.g. Öberg et al. 2011). Hydrogenated CO is overproduced, although not to such a high extent as in KS2013. It should be noted that it is methanol, not formaldehyde, that is its main form, which is a significant improvement thanks to the extended surface reaction set. $H_2CS$ is the main sulfur molecule in ice. $H_2S$, SO, $SO_2$, and OCS all have high and similar abundances. The major molecules for nitrogen are ammonia, $N_2$, HCN, HNCO, HNC, and $NH_2CHO$. Multi-atom organic species have high abundances, too. These are interesting results and major improvements over Kalvāns & Shmeld (2010, 2013). They pave the way for exploring the chemical role of the subsurface ice layers in a much better quality than before. A useful conclusion is that the reaction database is sufficient for mantle chemistry and does not need immediate extension. This is true at least for the major elements: C, N, and O.

In the above paragraph, the results with low diffusion energy barrier for surface reactions are considered (see Sect. 2.4. of S2010+). They can be directly compared to our previous results, however, the modified rate equation approach is more realistic. Simulation with the modified rate equations approach produces results fairly similar to those described above (Table 2). High diffusion energy ($E_b$ = 0.77 $E_D$) for surface species results in severe underproduction of $CO_2$ (Ruffle & Herbst 2001), while methanol is seriously overproduced. One can conclude that the modified rate equation approach is the best option, at least, when the model initial set-up is used, with parameters derived from KS2013 and given in this paper.

The presented results on ice composition are mostly the product of the well-developed "ALCHEMIC" model and OSU reaction database. The introduction of mantle chemistry is beneficial because it enhances the abundance of CO and $CO_2$ and helps to diminish the overproduction of methanol.

# 4. Conclusions

The exact values of the various efficiency parameters used in astrochemical modeling are often poorly known. The efficiency of all phase changes (accretion, desorption, ice re-cycling) as well as the photodissociation yield of ice species can be significantly altered within the frame of current knowledge. For example, there is a striking difference when the accretion rate is calculated either as outlined in Willacy & Williams (1993) or S2010+. Other parameters, such as integration time, temperature of gas and dust, cloud density, and extinction of the interstellar radiation field can be adjusted, too. The surface chemistry can be adjusted by changing the approach of doing calculations (simple rate equation, modified rate equations, slow diffusion, Monte Carlo random walk), or editing the reaction list.

The parameters in a model can be adjusted for the simulation results to fit the observational data. In such a way, a single model can account for a variety of cloud and protostar observations. This blessing comes also with the curse that a weak and incomplete model can be adjusted to produce seemingly correct results for a particular case. In many cases this can be done by adding or removing specific chemical reactions that are necessary for achieving the desired result. One has to be very careful when drawing conclusions from the results of models modified in a specific way.

Model results (calculated abundances) can be heavily dependent on the rates of a few phase-change processes and chemical reactions for a few major molecules. This was countered in two ways. First, a wider variety of phase change mechanisms (more than one mechanism for each transition, if physically feasible) in addition to conservative desorption yields were used. Second, the interpretation of results for the solid-phase species was done by drawing general trends. These include abundance ratios for elements in different phases. Especially, this regards the content of chemically bound hydrogen and deuterium.